# Detecting Suspicious Commenter Mob Behaviors on YouTube Using Graph2Vec

Shadi Shajari, Mustafa Alassad, Nitin Agarwal

COSMOS Research Center
University of Arkansas at Little Rock, USA
`{sshajari, mmalassad, nxagarwal}@ualr.edu`

**Abstract.** YouTube, a widely popular online platform, has transformed the dynamics of content consumption and interaction for users worldwide. With its extensive range of content creators and viewers, YouTube serves as a hub for video sharing, entertainment, and information dissemination. However, the exponential growth of users and their active engagement on the platform has raised concerns regarding suspicious commenter behaviors, particularly in the comment section. This paper presents a social network analysis-based methodology for detecting suspicious commenter mob-like behaviors among YouTube channels and the similarities therein. The method aims to characterize channels based on the level of such behavior and identify common patterns across them. To evaluate the effectiveness of the proposed model, we conducted an analysis of 20 YouTube channels, consisting of 7,782 videos, 294,199 commenters, and 596,982 comments. These channels were specifically selected for propagating false views about the U.S. Military. The analysis revealed significant similarities among the channels, shedding light on the prevalence of suspicious commenter behavior. By understanding these similarities, we contribute to a better understanding of the dynamics of suspicious behavior on YouTube channels, which can inform strategies for addressing and mitigating such behavior.

**Keywords:** Social Network Analysis, YouTube, Commenter Network Analysis, Graph2vec, UMAP, Suspicious Behaviors.

## 1    Introduction

YouTube, one of the largest video-sharing platforms in the world, has transformed the way people consume and engage with online content. With its vast user base and diverse range of videos, YouTube provides unrestricted space for creators and viewers to connect and interact. However, one notable issue is the presence of commenter mobs in the comment section. These are groups of commenters who comment together to boost the engagement of a video or multiple videos with comments from one or different channels. This behavior negatively affects user interactions and the overall authenticity of YouTube; where this behavior will disrupt the intended purpose of fostering meaningful dialogue and community engagement and poses significant challenges to user safety and platform integrity.

To address the problem of identification of commenter mobs and to better understand such problematic behavior; this paper presents a social network analysis-based methodology analysis of 20 YouTube channels that spread false information about the U.S. Military with the aim of identifying similarities between them. To achieve this goal, we employed advanced techniques such as Graph2Vec [1], which is a powerful tool for graph representation learning, Uniform Manifold Approximation and Projection (UMAP) for dimensionality reduction [2], and clustering methods, such as K-means and Hierarchical clustering [3]. Furthermore, we conducted both qualitative and quantitative analyses to assess the quality of the identified clusters. The analysis revealed significant similarities among the channels, shedding light on the prevalence of suspicious commenter behavior.

The remaining sections of the paper are structured as follows: In Section 2, we provide a comprehensive overview of suspicious behavior on YouTube, along with a discussion on Graph2vec, UMAP, and the current understanding of the topic. Section 3 outlines the data collection methods employed, highlighting the techniques and tools utilized. Section 4 describes the methodology utilized in the study, which combines Graph2vec, UMAP, K-means, and hierarchical clustering methods. Section 5 presents the study's findings, including an in-depth analysis of the data, commenter behavior, and analysis of YouTube channels. Finally, Section 6 concludes the paper and offers recommendations for future research.

## 2     Literature review

We classify the related works into three main sections, namely, Graph2Vec, UMAP and Suspicious Behavior on YouTube to cover the presented research topics.

### 2.1     Graph2Vec

Graph2Vec method is utilized to capture the graphs' attributes and structural information, representing unique labels to the fixed-length vectors using the Weisfeiler-leman Subtree kernel algorithm [1]. Originally, Narayanan et al. (2017) applied Graph2Vec to the context of social networks, where the authors captured information related to individuals' centrality values, neighborhood, and other structural details using a deep learning framework that utilizes Graph2Vec in a social network [1].

Yanardag et al. (2015) incorporated comparable techniques using the Wisfeiler-Lehman Subtree kernel, Grapghlet Kernels, and Shortest-Path graph kernels to capture structural information related to the large-scale graphs' similarities in the social network domain [4]. Ribeiro et al. (2017) introduced the struc2Vec framework to employ nodes' structural identity and neighborhood information to improve the representation of the nodes in graphs and improve performance on classification tasks [5]. In addition, Donnat et al. (2018) also applied the "GraphWave", a spectral graph wavelets method that represents each node's network neighborhood via a low-dimensional embedding [6]. Furthermore, Zeng et al. (2019) proposed GrahpSAINT method using a graph sampling based training method, where the authors utilized deep Graph Convolutional Networks (GCNs) on various social networks like Reddit [7].

Likewise, Varlamis et al. (2022), applied GCNs method to improve the graph and nodes presentations to the task of detecting fake news, fake news accounts, and rumors that spread in social networks in Twitter [8]. Also, Cekinel et al. (2022) applied subgraph embedding and graph sequence mining to predict the Turkish news events skeleton in the form of a subgraph and sequence rules in Turkish newspaper, such as Dünya Gazetesi's and CC-News datasets [9].

## 2.2    Uniform Manifold Approximation and Projection (UMAP)

Colleagues realized numerous research methods and applications for UMAP. Just to name a few, these methods have showcased UMAP as a powerful dimension reduction technique capable of visualizing similarity in high dimensional datasets [2], performing general non-linear dimension reduction [10], and an approach to capture the global structure of high-dimensional data accurately as presented in [11]. Furthermore, UMAP is constructed from a theoretical framework based on Riemannian geometry and algebraic topology, where UMAP has no computational restrictions on the embedding dimension, making it viable as a general purpose dimension reduction technique for machine learning, as explained by [12].

Likewise, Ghojogh et al. (2021) surveyed the applications of the UMAP algorithm, where the authors explored the UMAP approach's probabilities of the neighborhood in the input and embedding spaces, optimization of the cost functions, training algorithms, derivation of gradients, and supervised and semi-supervised embedding applications by UMAP [2]. For this study, we implemented Graph2Vec and UMAP to reduce the high dimensionality, linearize the complexity in computations, and capture the structural details of the commenters on YouTube channels.

## 2.3    Suspicious Behavior on YouTube

Many colleagues have proposed various studies related to suspicious behaviors on YouTube in this domain. Alassad et al. [13] proposed a two-level decomposition method to identify focal structure sets on YouTube responsible for disseminating conspiracy theories. Likewise, Alassad et al. [14] applied the bi-level max-max optimization approach to study individuals' local and global impacts on social media. In other words, the study explores structural information on key structures that have the power and resources to mobilize crowds and control the flow of information.

Furthermore, Hussain et al. [15] proposed the engagement scores approach to identify inorganic behaviors on YouTube networks. Likewise, Kirdemir et al. [16] proposed an unsupervised methodology that combines multiple layers of analysis to explore coordinated inauthentic behavior assessment on YouTube. This research addresses the knowledge gap using a systematic approach to explore structural information at individual and multi-channel levels.

It is important to mention that the choice of datasets depends on the research questions, application domain, and availability of the data; this research incorporates the Graph2Vec approach and the Wisfeiler-Lehamn Subtree Kernel algorithm to extract structural suspicious information on YouTube channels.

## 3    Data Collection

To collect data on the U.S. Military-related YouTube channels, we utilized a Python-based multi-threaded script [17]. Subject matter experts helped us select a list of 20 channels that promoted false views of the U.S. military. Using the YouTube Data API, we collected extensive data on these channels, including the number of videos, comments, details of the comments, and IDs of the commenters. The data we collected includes 20 channels, which collectively contain 7,782 videos, 596,982 comments, and involve 294,199 commenters.

## 4    Methodology

This section presents the organization of the proposed model; initially, we outline the procedure for creating a co-commenter network in a suitable structure for our model (Section 4.1). Following that, we delve into the concepts of Graph2Vec and UMAP (Sections 4.2 and 4.3). Subsequently, our model employs the employment of K-means and hierarchical clustering methods [3]. Lastly, we used the resulting embeddings to cluster the co-commenter networks based on their similarity.

### 4.1    Creating Co-commenter Network

This network for each of the 20 channels was constructed based on the procedure and the predetermined threshold explained in [18]. Likewise, the constructed co-commenter network comprises edges between commenters who have left comments on the same video in one or several channels according to the procedure structured in [16]. The weight of these edges is determined by the number of shared videos on which they have commented. To utilize the co-commenter network in the model, we exported it as a gexf file, which is a format based on XML that facilitates the storage of a singular graph structure [19]. The next section describes the implementation of Graph2Vec.

### 4.2    Graph2Vec

In this section, we describe our methodology for analyzing the co-commenter networks using the Graph2Vec approach. Graph2Vec represents graphs in a vector space to capture their unique patterns. To apply Graph2Vec to the co-commenter networks, we utilized the unique procedure proposed in [19]; where the method employs the Weisfeiler-Leman Algorithm (WLA) [10]. The solution procedure utilizes iterative labeling of all nodes based on local neighborhood structures to capture details of complex patterns. Furthermore, the key parameters were carefully selected, including a WLA iteration of 2 to identify structural patterns beyond immediate neighborhoods, a dimensionality of 128 for balanced information capture, a learning rate of 0.025 for stable optimization, and a minimum word count of 5 for semantic understanding. Using these default parameters, the fit method of the Graph2Vec object trained the model on multiple co-commenter networks.

Moreover, the solution procedure involved extracting feature representations using the Weisfeiler-Leman algorithm, generating hashed neighborhood labels. These features were transformed into TaggedDocument objects and utilized the Doc2Vec model to learn continuous vector representations of the co-commenter networks. To recap this section, the methodology combines Graph2Vec and textual analysis to extract meaningful insights from co-commenter networks, with the selection of parameters ensuring optimal performance and effective representation of the networks.

### 4.3    UMAP

UMAP is utilized to reduce the dimensionality of the Graph2Vec embeddings and visualize the extracted patterns effectively [10]. We carefully selected all default parameters to ensure optimality in the performance and meaningful visualizations. The adapted procedure includes the number of neighbors, set to 5, for balancing the local structure preservation, the minimum distance between points, set to 0.1, for appropriate spread and separation, and the number of components, set to 4, for capturing significant structural information while facilitating the visualization. In addition, the standard Euclidean distance metric was utilized to observe the performance of the UMAP algorithm successfully reduced the dimensionality, preserved the global structure, and facilitated accurate visualization and clustering of the co-commenter networks.

## 5    Result

In this section, we present the results of our clustering analyses using both the K-means and hierarchical clustering methods [20]. To further assess the quality of the identified clusters, we conducted both qualitative and quantitative cluster quality analysis [21].

### 5.1    K-Means and Hierarchical Clustering

During the analysis, two distinct clusters were identified by both techniques as shown in Figure 1. The optimal number of clusters for K-means was determined using the silhouette score, which measures the similarity among channels within each cluster as described in [22]. Similarly, for hierarchical clustering as presented in [23], the single linkage method was employed to create a dendrogram to present the closeness of each channel to others. The cut-tree method was utilized to determine the optimal number of clusters and separate the channels based on their similarity.

In Figure 1, the image on the right side represents K-means clustering results; the green group corresponds to cluster 0, while the red group corresponds to cluster 1. Likewise, these visualizations, such as scatter plots [22], revealed well-separated clusters, indicating the effective classification of data points within each cluster. Likewise, this demonstrates the quality and accuracy of our clustering approach accurately grouping similar channels together. Furthermore, Figure 1 left side, demonstrates the results obtained from hierarchical clustering, where the channel index refers to the unique channel ID assigned to each channel (graph) in our dataset.

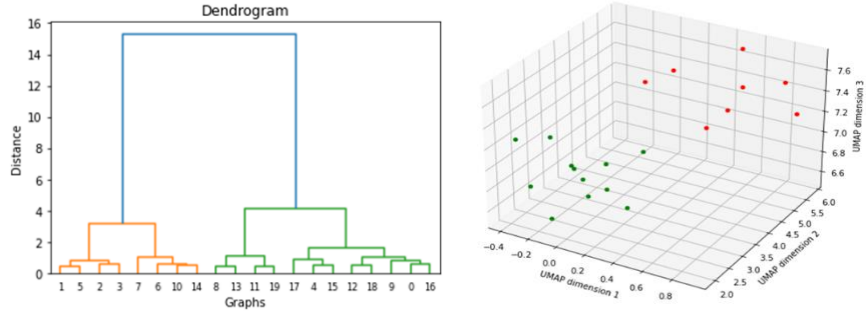

Figure 1. Channel categorization using hierarchical clustering (left) and k-means(right).

## 5.2 Qualitative and Quantitative Cluster Quality Analysis

The qualitative analysis required visually examining the scatter plots to observe the clear distinction and unity among the clusters, as illustrated in Figure 1, and the size of the co-commenter network for each channel. It also involved assessing the similarity of the video content shared among the channels within each cluster. Meanwhile, the quantitative analysis concerned calculating relevant metrics such as the cophenetic correlation coefficient [24], and the Davies-Bouldin index [25]. These measures provide objective indicators of how well the clustering algorithm successfully grouped similar channels together and separated dissimilar ones. The qualitative and quantitative analysis results provide strong evidence supporting our clustering approach's effectiveness and the identified clusters' reliability as explained below.

During the quantitative cluster quality analysis, we assessed the performance of the clustering algorithm using two measures. The first measure, the cophenetic correlation coefficient, yielded a score of 0.82. A higher score indicates better separation between the clusters, indicating that the algorithm successfully grouped similar channels together. The second measure, the Davies-Bouldin index, resulted in a score of 0.44. A lower score indicates better separation between the clusters, suggesting that the algorithm effectively differentiated dissimilar channels.

Furthermore, to analyze the distinctive features of the identified clusters, another important aspect of our research involved determining the most suspicious channels within a cluster. To accomplish this, we investigated the presence of cliques within the commenter network and specifically focused on cliques with at least five members. Cliques raised suspicions because they displayed strong connections and coordinated actions among commenters of a channel. By examining these suspicious commenter cliques, we could investigate individual channels to determine if they are involved in suspicious or coordinated commenters' mob-like activities. In our analysis, we compared the two clusters, Cluster 0, and Cluster 1, to understand the presence of suspicious channels within each cluster. Cluster 0 emerged as the cluster with channels comprising of commenters with the highest number of cliques containing at least five members, indicating a significant concentration of suspicious channels. Within Cluster 0, we identified Channels "Armageddon TV", "USA Military Channel", and "USA Military

Channel 2" as the top three most suspicious channels, with cliques numbering 12,241, 9,246 and 782 respectively.

These findings highlight the potential for coordinated activities and interconnectedness within Cluster 0. In contrast, Cluster 1 exhibited a smaller number of cliques, with Channels "US Military System", "Brittany Lewis", and "U.S. Military Technology" being the top three most suspicious channels, possessing 26, 24, and 20 cliques respectively. Although the number of cliques in Cluster 1 is relatively lower than in Cluster 0, it still raises concerns regarding potential suspicious activities within this cluster. In the context of qualitative analysis, we observed distinctive traits in the identified clusters by closely examining YouTube channels and their content.

Cluster 0 consisted of larger channels claiming official sources, focusing more on US military news compared to news from other countries. These insights provide valuable information about the composition and unique characteristics of channels in each cluster, helping us understand their differences and similarities. Figure 2 presents the example videos representing the channels within cluster 0.

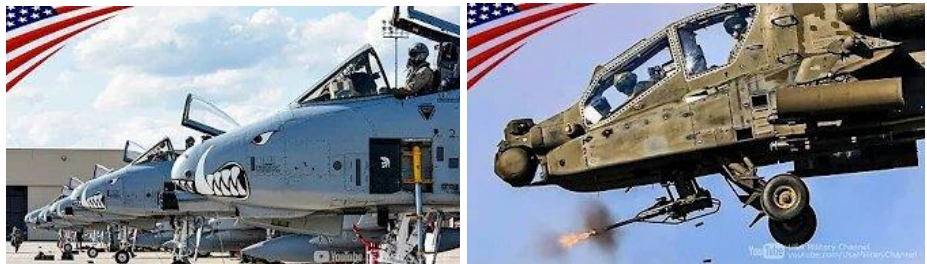

Figure 2. Thumbnail images of a video from USA Military channel (left) and USA Military channel 2 (right).

On the other hand, cluster 1 included relatively smaller channels associated with non-official news sources and personal accounts. These channels mainly featured videos about the US military and covered topics related to countries like Ukraine, Russia and other countries. This highlights the diversity of content and perspectives available on YouTube, ranging from individual viewpoints to alternative sources of news. Figure 3 illustrates the sample videos that depict the channels belonging to cluster 1.

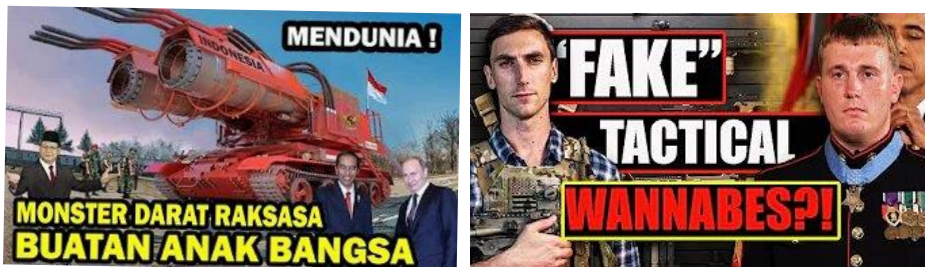

Figure 3. Thumbnail images of a video from ACTUAL NEWS TODAY (left) and NICKY MGTV (right).

Overall, the combination of cluster analysis and clique examination enables a comprehensive approach to studying suspicious channels within a given network. This methodology allows us to identify clusters with a high concentration of cliques and subsequently investigate the channels within those clusters that exhibit the most suspicious characteristics.

## 6    Conclusion and Future Works

In this study, we propose a methodology for identifying and detecting suspicious commenter behaviors in co-commenter networks, also known as "commenter mobs", for 20 YouTube channels that disseminate false information about the U.S. Our approach involves utilizing Graph2Vec to learn embeddings for the networks, enabling accurate clustering based on similarity. Additionally, we incorporated dimensionality reduction using UMAP to enhance the clustering process. Through qualitative and quantitative analysis, we comprehensively evaluated the clustering algorithm, validating the quality of the resulting clusters. Furthermore, we observed distinct characteristics of the clique structures between the two clusters (Cluster 0 and Cluster 1). In Cluster 0, we identified a higher number of cliques with at least five members, suggesting a higher degree of coordinated activities and potential suspicious behavior among the commenters of channels within this cluster. On the other hand, Cluster 1 exhibited a smaller number of such cliques, indicating a relatively lower level of coordinated activity. These findings underscore the significance of leveraging advanced techniques for identifying suspicious mob-like behaviors in YouTube co-commenter networks. Although this research is conducted on the YouTube platform, it is easily generalizable to analyze commenter behaviors on other social media platforms.

For future work, there are several avenues to explore based on the findings of our study. First, further investigation can be conducted to explore the impact of different graph embedding techniques in comparison to Graph2Vec. This would allow for a more comprehensive understanding of the strengths and limitations of various embedding approaches in the context of co-commenter networks. Additionally, the scalability and performance of the clustering algorithm can be assessed by applying it to larger datasets with a higher number of YouTube channels. This would help determine the algorithm's effectiveness in handling larger and more complex networks, providing insights into its applicability to real-world scenarios.

## Acknowledgment

This research is funded in part by the U.S. National Science Foundation (OIA-1946391, OIA-1920920, IIS-1636933, ACI-1429160, and IIS-1110868), U.S. Office of the Under Secretary of Defense for Research and Engineering (FA9550-22-1-0332), U.S. Office of Naval Research (N00014-10-1-0091, N00014-14-1-0489, N00014-15-P-1187, N00014-16-1-2016, N00014-16-1-2412, N00014-17-1-2675, N00014-17-1-2605, N68335-19-C-0359, N00014-19-1-2336, N68335-20-C-0540, N00014-21-1-2121, N00014-21-1-2765, N00014-22-1-2318), U.S. Air

Force Research Laboratory, U.S. Army Research Office (W911NF-20-1-0262, W911NF-16-1-0189, W911NF-23-1-0011), U.S. Defense Advanced Research Projects Agency (W31P4Q-17-C-0059), Arkansas Research Alliance, the JerryL. Maulden/Entergy Endowment at the University of Arkansas at Little Rock, and the Australian Department of Defense Strategic Policy Grants Program (SPGP) (award number: 2020-106-094). Any opinions, findings, and conclusions or recommendations expressed in this material are those of the authors and do not necessarily reflect the views of the funding organizations. The researchers gratefully acknowledge the support.